\documentclass[aps,pra,twocolumn,letterpaper, floatfix, superscriptaddress,amsmath,amssymb]{revtex4-2}
\usepackage{amsmath,amssymb, graphicx}
\usepackage{bm,times, color}
\usepackage[utf8]{inputenc}
\usepackage[colorlinks,breaklinks]{hyperref}
\usepackage[svgnames]{xcolor}
\hypersetup{linkcolor=DarkBlue, citecolor=DarkBlue, filecolor=black, urlcolor=DarkBlue}
\usepackage{comment, mathrsfs,slashed}
\usepackage{amsfonts}
\usepackage{bbm}
\usepackage[normalem]{ulem}
\def \p{\partial}
\def \dag{\dagger}
\def \mb{\mathbf}

\def \lan{\langle}
\def \ran{\rangle}

    \makeatletter
\def\@fnsymbol#1{\ensuremath{\ifcase#1\or \dagger\or \ddagger\or
   \mathsection\or \mathparagraph\or \|\or **\or \dagger\dagger
   \or \ddagger\ddagger \else\@ctrerr\fi}}
    \makeatother

\begin{document}
\title{Precise correspondence between the $p$-wave chiral superfluid and the spinless bosonic superfluid in the lowest Landau level}
\author{Wei-Han Hsiao}
\altaffiliation{Ph.D. in Physics, The University of Chicago}
\noaffiliation{}

\date{Jan 8th, 2025}

\begin{abstract}
We establish a precise formal correspondence between a spinless $p$-wave chiral fermionic superfluid and a spinless bosonic superfluid in a strong magnetic field by correctly accounting for superfluid vorticity in the effective theory of the former. In the literature, this vorticity term incompletely manifests as the magnetic field. This paper demonstrates this substitution can be understood as a truncation within the relevant expansion scheme, accompanied by field redefinitions. The components discarded in this truncation are critical for restoring the Berry phase term in the effective theory, encapsulating both systems in the same master Lagrangian. Beyond clarifying the structure of the Berry phase, this formalism allows for solving the bosonic system in the lowest Landau level (LLL) by analogy. Specifically, we show that, in the linear regime, the Maxwell equations governing these systems are identical when the vortex crystal is reformulated using an auxiliary electromagnetic field. This approach offers a unified perspective on these systems and yields solutions that are rotationally covariant, gauge invariants, and physically interpretable.
\end{abstract}

\maketitle
\section{Introduction}
The phenomenon of superfluidity has been a wonderland of inspiration since its discovery in the early 20th century. In this realm, microscopic particles, regardless of their statistical properties, can exhibit their quantum characteristics through macroscopic wave functions. This results in fascinating phenomena such as fountain effect and viscosity-less flow in superfluid ${}^4$He. In the presence of electromagnetic interactions, charged superfluids become superconductors via the Higgs mechanism, exhibiting effects such as the Meissner effect and the Josephson effect \cite{schrieffer1999theory}. Even more compelling, in the presence of rotation, superfluids can host exotic many-body states, such as vortex crystal \cite{RevModPhys.59.87} and bosonic version of the fractional quantum Hall effect \cite{Fetter2008, doi:10.1080/00018730802564122}.

The standard paradigm explaining superfluid physics is based on the models of Bogoliubov, and Bardeen, Cooper, and Schrieffer (BCS) for bosonic and fermionic systems, respectively \cite{landau1980course, schrieffer1999theory, Abrikosov:107441}. The accompanying field-theory inclined approaches, the Gross-Pitaevskii equation and the Landau-Ginzburg-Gor'kov equation, are still actively employed in research. Relatively recently, physicists have begun reexamining the large-scale physics from the perspective of effective field theories (EFT) \cite{weinberg1995quantum}. In this framework, the Lagrangian is constructed based on governing symmetries and a gradient expansion scheme that organizes the relevance of plausible physics according to the exponent of the spacetime derivative. This approach has been applied to $s$-wave fermionic superfluids, bosonic superfluid, and $p$-wave fermionic chiral superfluids that respect a generalized Galilean symmetry \cite{SON2006197, PhysRevB.89.174507, PhysRevB.91.064508, PhysRevLett.122.235301, Baiguera2024}, referred to as non-relativistic spacetime diffeomorphism.

Among these examples, this paper focuses on the $p$-wave fermionic chiral superfluid and the bosonic superfluid in a background magnetic field in (2+1) spacetime dimensions. Besides the non-relativistic diffeomorphism, both break parity ($\mathcal P$) and time-reversal ($\mathcal T$) symmetries, making it tempting to formulate them in a coherent fashion. In particular, as shown in Ref.\cite{PhysRevLett.122.235301}, the latter includes a Berry phase term, originating from the vorticity of the superfluid flow, which contributes to the next-to-leading order effective Lagrangian in the limit of the LLL projection. Clarifying the corresponding structure in the former serves as a step toward this goal.

We will demonstrate that formal correspondences between these systems exist at various stages. First, we will offer a derivation for the $p$-wave chiral superfluid in which the target Berry phase term is made manifest. In the dual photon representation, the manifestation identifies its superfluid sector and that of the bosonic superfluid. We then reformulate the vortex crystal sector of the bosonic superfluid using the field strength of an auxiliary U(1) gauge field, where the Magnus force becomes a mixed Chern-Simons term. Consequently, the resulting linearized Maxwell equations for the superfluid sector are identical to those from the $p$-wave chiral superfluid, allowing the solutions to be extracted straightforwardly by analogy, leading to much simpler routes for calculating collective excitations and linear responses. As an illustration, in the limit of long wavelength and LLL projection, the electromagnetic response of a bosonic $s$-wave superfluid is encoded in the following simple Lagrangian:
\begin{align}
\label{mainLeff}\mathscr L_{\rm eff} = &\frac{n_0}{2B_0}\epsilon^{\mu\nu\rho}A_{\mu}\p_{\nu}A_{\rho} +  \frac{(\mathsf g-2)n_0}{4mB_0}B^2 \notag\\
&+ \frac{n_0(\mathsf g-4)}{4B_0^2}B\nabla\cdot\mb E + \cdots.
\end{align}
In addition to bridging two plausibly similar systems, our approach offers several advantages and complements existing knowledge in a few key ways: It is manifestly gauge-invariant and rotationally covariant, with no need to specify a gauge choice or spatial direction. This approach reduces the superficially $4\times4$ matrix system \cite{PhysRevLett.122.235301} to elementary Gauss' law and Amp\`{e}re's law, making the solutions more comprehensible and their long-wavelength expansions physically interpretable. Last but not least, the introduction of the auxiliary electromagnetic field allows us to view superfluid-vortex crystal dynamics as a gauge theory \cite{10.21468/SciPostPhys.14.5.102, DAYI2024129285}, providing an alternative toolkit for investigating or reexamining the physical properties of this and related systems.

The rest of the paper is organized as follows. In Sec.\ref{pwave}, we provide a minimalistic yet self-contained review of the effective theory of superfluids using non-relativistic diffeomorphism. Following this, we employ a master Lagrangian~\eqref{Lmaster} and elaborate on the necessary steps to produce the results from Ref.\cite{PhysRevB.89.174507} and Ref.\cite{PhysRevB.91.064508}. The key discrepancy, which leads to the introduction of the desired Berry phase correction not previously accounted for in the literature, will be highlighted and rationalized. Sec.\ref{swave} begins with deriving the superfluid sector of the bosonic superfluid in the LLL using~\eqref{Lmaster} with the proper inclusion of the $\mathsf g$ factor. In the subsequent linear response calculation, we demonstrate the deduction of the solutions from Sec.\ref{pwave} based on mathematical analogy. Concluding remarks and futures directions are presented in Sec.\ref{conclusion}.

\section{The effective theory of a $p$-wave chiral superfluid}\label{pwave}
\subsection{Convention and notations}
As a significant portion of the derivation involves curved space, let us first specify the conventions and notations employed throughout the main text. 

We will work in a (2+1) dimensional spacetime. The spacetime indices $0, 1$, and $2$ are denoted by Greek letters $\mu, \nu, \rho, \lambda,$ etc. The time dimension $t$ is considered absolute, and its index is simply $0$. The spatial dimensions $1$ and $2$ can be curved, and are indexed by Latin letters $i, j, k, l,$ etc. Thus, the components of the spacetime coordinate are written as $x^{\mu} = (t, \mb x) = (t, x^i)$. Additionally, the local orthonormal coordinates are indexed by capital Latin letters $A, B, C,$ etc. 

On the spacetime, the curvature is described with the metric tensor $g_{ij}(t, \mb x)$, which is assumed to have an inverse $g^{ij}$. In the local orthonormal frames, the metrics are simply $\delta_{AB}$. The volume measure is $d^3x\, \sqrt{g}$, where $g = \det g_{ij}$. In this context, the metric tensor serves as a prescribed background probe field that facilitates the computation of gravitational responses and renders the effective action invariant under diffeomorphism. It is analogous to the background U(1) electromagnetic potential $A_{\mu}$, which facilitates evaluating charge current and manifests gauge symmetry.

The Levi-Civita symbol is specified as $\epsilon^{012} = 1$ and $\epsilon^{ij} = \epsilon^{0ij}$. The Levi-Civita tensor is denoted as the curly version $\varepsilon^{\mu\nu\rho} = \epsilon^{\mu\nu\rho}/\sqrt{g}$. $\epsilon^{AB}$ is numerically defined the same as $\epsilon^{ij}$.

Lastly, for calculation in flat spaces, the spatial indices are lowered and raised using the Kronecker delta, and we will not rigorously distinguish contravariant and covariant indices. In these instances, we will also exploit the standard notations for curl and divergence from vector calculus: $\epsilon^{ij}\p_iv_j = \nabla\times\mb v$ and $\p_i v^i = \nabla\cdot\mb v$.
\subsection{Building blocks of the effective theory}
The general idea of constructing an EFT based on symmetry principle is as follows: We first consider a generic microscopic model that contains the necessary ingredients to produce the desired macroscopic physics. We then identify the symmetries obeyed by this model and how each field transforms under these symmetries. The relative magnitudes of the fields are determined by a rule for power counting, which assigns a power of momentum to each field. In the limit of long wavelengths, combinations of fields with high momentum powers are truncated. This way, the dominant contribution to the low-energy physics can be systematically extracted. \cite{SON2006197}. 

Let us consider the following non-interacting model in a (2+1) dimensional spacetime as a brief example:
\begin{align}
\label{Lnon}\mathscr L_{\rm non} = \frac{i}{2}\psi^{\dag}\overset{\leftrightarrow}{D}_t \psi - \frac{g^{ij}}{2m}D_i\psi^{\dag}D_j\psi,
\end{align}
where $\psi$ is the particle field, and $D_{\mu} = \partial_{\mu} - i\varsigma A_{\mu}$ is the covariant derivative. Here, $\varsigma = \pm1$ accommodates different charge conventions in the literature. The relevant symmetries include the U(1) symmetry of total particle number $\psi \to e^{i\alpha}\psi, A_{\mu}\to A_{\mu}-\p_{\mu}\alpha$ and the non-relativistic diffeomorphism:
\begin{align}
\label{transform}  x^{\mu} \to x^{\mu} + \xi^{\mu},\ \xi^{\mu} = (0, \xi^i(t, \mb x))
\end{align}
It can be directly verified that the Lagrangian transforms as a scalar,  $\delta\mathscr L_{\rm non} = -\xi^k\p_k\mathscr L$, when $\psi$, $g_{ij}$ and $A_{\mu}$ transform as follows:
\begin{subequations}
\begin{align}
& \delta\psi = -\xi^k\p_k\psi\\
& \delta g_{ij} = -\xi^k\p_kg_{ij} - g_{kj}\p_i\xi^k - g_{ik}\p_j\xi^k\\
\label{dA0}& \delta A_0 = -\xi^k\p_k A_0 - A_k\dot{\xi}^k \\
\label{dAi}& \delta A_i = - \xi^k\p_k A_i  - A_k\p_i \xi^k  -\varsigma mg_{ik}\dot{\xi}^k,
\end{align}
\end{subequations}
To apply this formalism to superfluids, note that at low energies, the light fields are the gradient of the phase $\psi = |\psi|e^{-i\theta}$, $\p_{\mu}\theta$, as well as the background parameters $A_{\mu}$ and $g_{ij}$. By virtue of the symmetry principle, it is well-understood that the following quantity is of order one and transforms as a scalar under diffeomorphism \cite{SON2006197, doi:10.1142/S0217984989001400}: 
\begin{align}
X = D_t\theta -\frac{g^{ij}}{2m}D_i\theta D_j \theta, D_{\mu}\theta = \p_{\mu}\theta +\varsigma A_{\mu}.
\end{align}
As a result, the leading order Lagrangian of a plain vanilla superfluid, such as a $s$-wave BCS superfluid, is a function of $X$. 

To extend the above to capture the physics of chiral superfluids, we note that the Cooper pairs in a chiral superfluid carry orbital angular momentum, implying a geometric response in the low-energy effective action similar to the effect of microscopic spins \cite{PhysRevB.110.024515}. To illustrate the consequence, consider the ground state condensate of the form
\begin{align}
\Delta_{\mb p} = (p^x + \sigma i p^y)\lan\psi_{-\mb p}\psi_{\mb p}\ran, \sigma = \pm 1.
\end{align}
Under a local SO(2) rotation, $p^A \to p^A -\phi \epsilon^{AB} p^B$, the ground state acquires a phase shift $\delta\Delta_{\mb p} = i\sigma\phi \Delta_{\mb p}$, which must be compensated by the phase of $\psi$ to maintain ground state invariance. The corresponding transformation introduces a term $\frac{\sigma}{2}\p_{\mu}\phi$ in the covariant derivative, and a connection $\omega_{\mu}\to \omega_{\mu} + \p_{\mu}\phi$ should be introduced to eliminate this gradient. Consequently, the covariant derivative $D_{\mu}\theta$ acquires a spin connection $\omega_{\mu}$:
\begin{align}
D_{\mu}\theta \to \p_{\mu}\theta +\varsigma A_{\mu} - s\omega_{\mu}, s = \pm \frac{1}{2}.
\end{align}
The construction of $\omega_{\mu}$ is more involved but follows standard procedures \cite{Son:2013rqa, PhysRevX.7.041032}. Nevertheless, the non-relativistic diffeomorphism introduces a nuance to $\omega_0$ to ensure that $\omega_{\mu}$ transforms as a one-form. Recall that in ordinary differential geometry, we introduce an orthonormal spatial vielbein $e^{A}\!_j$ satisfying the properties 
\begin{align}
g_{ij} = \delta_{AB}e^{A}\!_i e^{B}\!_j,\ \epsilon_{AB}e^{A}\!_ie^{B}\!_j = \varepsilon_{ij}.
\end{align}
The local SO(2) rotation modifies it by $e^{A}\!_j \to e^A\!_j + \phi \epsilon^{A}\!_Be^B\!_j$, and the diffeomorphism transforms its $j$ index as a one-form. The spin connection is defined as the exterior derivative of the vielbein. This conventional connection, however, fails to transform correctly due to the time derivative in $\omega_0$. To resolve this, we postulate a velocity field $v^{\mu} = (1, v^i)$, which transforms as a contravariant vector under Eq.~\eqref{transform}:
\begin{align}
\delta v^{\mu} = -\xi^{\lambda}\p_{\lambda}v^{\mu} + v^{\lambda}\p_{\lambda}\xi^{\mu}.
\end{align}
Appending its vorticity field $\varepsilon^{ij}\p_iv_j$ (with $v_i = g_{ij}v^j$) to the temporal component, the following improved connection $\omega_{\mu}$ transforms as a one-form.
\begin{subequations}
\begin{align}
\label{omegaZero}& \omega_0 = \frac{1}{2}\left( \epsilon^{ab}e^{aj}\p_t e^b_j + \varepsilon^{ij}\p_i v_j \right)\\
\label{omegaI}& \omega_i = \frac{1}{2}\left( \epsilon^{ab}e^{aj}\p_i e^b_j - \varepsilon^{kl}\p_kg_{il}\right).
\end{align}
\end{subequations}
Moreover, the velocity field facilitates the improved U(1) connection:
\begin{subequations}
\begin{align}
\label{tildeA0}& \tilde A_0 = A_0 - \varsigma\frac{m}{2}g_{ij}v^iv^j\\
& \tilde A_i = A_i + \varsigma m v_i
\end{align}
\end{subequations}
Based on Eqs~\eqref{dA0} and~\eqref{dAi}, $\tilde A_{\mu}$ also transforms as a one-form. Note that we have adopted a more bottom-up approach to define these foundational ingredients to make the review section compact. Alternatively, one could adopt a more top-down approach by conceiving tensors in a Newton-Cartan spacetime \cite{Son:2013rqa, PhysRevB.91.064508}, where the geometric objects $\omega_{\mu}$ and $v^{\mu}$ appear more naturally by construction and coincide with our definition in a global time frame.

We have gathered the necessary covariant quantities $\p_{\mu}\theta$, $v^{\mu}$, $\tilde A_{\mu}$ and $\omega_{\mu}$ for a model of the chiral $p$-wave superfluid. Denote the particle number density, a scalar under diffeomorphism, by $\rho$. The following master Lagrangian satisfies all desired transformation properties and is consistent with existing superfluid models:
\begin{align}
\label{Lmaster}\mathscr L_{\rm M} = \rho v^{\mu}(\p_{\mu}\theta +\varsigma \tilde{A}_{\mu}-s\omega_{\mu}) - \epsilon(\rho),
\end{align}
where $\epsilon(\rho)$ denotes the energy density. Although the meaning of $v^{\mu}$ and $\rho$ might not have been clear when they were introduced, their physical significance becomes evident in~\eqref{Lmaster} as they couple to $A_{\mu}$ through the form $J^{\mu}A_{\mu}$. In this effective Lagrangian, $\rho, v^{\mu}$ and $\theta$ are all dynamical and need to be solved by the field equations of motion. We will soon examine example solutions.

Model \eqref{Lmaster} has been partially explored in the literature. Specifically, when $s = 0$ and $\varsigma = -1$, we can precisely derive the equivalence between the actions presented in Ref.\cite{PhysRevB.89.174507} and Ref.\cite{PhysRevB.91.064508} by solving $v^i$ in two ways. The steps are outlined in the main text and the appendix of Ref.\cite{PhysRevB.91.064508}. Nevertheless, neither Ref.\cite{PhysRevB.89.174507} nor Ref.\cite{PhysRevB.91.064508} demonstrates this equivalence for $s \neq 0$ using the same logical steps. More precisely, the generalization is argued, and the vorticity term in $\omega_0$ is identified with the magnetic field $B$ merely by the transformation properties under diffeomorphism. This approach would leave ambiguity, as the transformation remains unchanged if one shifts $\omega_0$ by a spacetime scalar: $\omega_0 \to \omega_0 + f$, where $\delta f = -\xi^k\p_kf$. We will resolve this ambiguity and provide a more comprehensive solution in the following subsection.

\subsection{The accurate vorticity term}
Due to the problem discussed in the last paragraph, the rest of this section will go beyond the existing literature to address the ambiguity. We will present two ways to eliminate the velocity field. The first method directly solves for it and consistently accounts for the boundary term contribution from the vorticity term. The second method employs particle-vortex duality, rephrasing it in terms of the dual photon field. Upon examining the final form of the effective Lagrangian, we will highlight the discrepancies from the EFTs in Refs.\cite{PhysRevB.89.174507} and \cite{PhysRevB.91.064508}, explaining how these are compatible across both solution schemes. For explicit comparison with the results in the literature, throughout this subsection, we set $\varsigma =-1$ .
\paragraph{Direct integration}
Varying $\int d^3x\sqrt{g}\mathscr L_{\rm M}$ with respect to $v^i$, we obtain:
\begin{align}
 v^i = -\frac{1}{m}g^{ij}(\p_j \theta - A_j -s\omega_j) + \frac{s}{2m}\varepsilon^{ij}\p_j \log \rho.
\end{align}
The velocity consists of a convection term from the conventional superfluid velocity and a boundary term, giving rise to the following vorticity:
\begin{align}
\label{rotv}\varepsilon^{ij}\p_i v_j = \frac{1}{m}(B + s\varepsilon^{ij}\p_i\omega_j) - \frac{s}{2m}\nabla_i\nabla^i \log\rho.
\end{align}
As is physically understood, when the velocity field is a pure potential flow, $\varepsilon^{ij}\p_iv_j = 0$. Eq.~\eqref{rotv} highlights the sources that break this condition, the well-known magnetic field and geometric curvature. The contribution from the fluctuation of particle density, or edge flow contribution, in Eq.\eqref{rotv} might seem peculiar. We note that it arises from defining the velocity to be directly proportional to the U(1) current $J^{\mu} = \rho v^{\mu}$. This edge flow term is often redefined separately as an additional contribution to the U(1) current. Let us write:
\begin{subequations}
\begin{align}
& v^i = - \frac{1}{m}g^{ij}\bar v_j + \frac{s}{2m}\varepsilon^{ij}\p_j \log \rho\\
\label{v0bar}& \p_t\theta - A_0 - s\omega_0 = \bar v_0 + \frac{s^2}{4m}\nabla_i\nabla^i \log \rho.
\end{align}
\end{subequations}
The pure convectional $\bar v_i$ would satisfy a more conventional vorticity condition $\varepsilon^{ij}\p_i \bar v_j \propto (B + s\varepsilon^{ij}\p_i\omega_j)$. Plugging the above solutions back, the Lagrangian then becomes:
\begin{align}
\label{Lfirstway} \rho\left( \bar v_0 - \frac{1}{2m}g^{ij}\bar v_i\bar v_j \right) + \frac{s^2}{8m}\rho\nabla_i\nabla^i\log\rho - \epsilon(\rho).
\end{align}
In this form, it is evident that~\eqref{Lfirstway} is not a Legendre transform of $\epsilon(\rho)$ with respect to $\rho$. Equivalently~\eqref{Lmaster} is not a function of $\left( \bar v_0 - \frac{1}{2m}g^{ij}\bar v_i\bar v_j \right)$, unless the Laplacian of the density profile is truncated. Additionally, compared to the temporal SO(2) connection defined in Refs.\cite{PhysRevB.89.174507,PhysRevB.91.064508}, $\bar v_0$ contains an extra Ricci scalar term $-\frac{s^2}{2m}\varepsilon^{ij}\p_i\omega_j$. Therefore, we may understand that, in this derivation, the model~\eqref{Lmaster} implies the leading-order Lagrangian in Ref.\cite{PhysRevB.89.174507}, up to order $\mathscr O(s)$.
\paragraph{Dual photon representation}
We will demonstrate this statement using the dual photon picture. Varying $\int d^3x\sqrt g\, \mathscr L_{\rm M}$ with respect to $\theta$, the equation of motion implies the conservation of U(1) charge: $g^{-1/2}\p_t(\sqrt g\, \rho) + \nabla_i(\rho v^i) = 0$. The standard solution involves introducing the dual photon, $a_{\mu}$, such that $(\rho, \rho v^i) = \varepsilon^{\mu\nu\rho}\p_{\nu}a_{\rho}$, or in terms of the field strength, $\rho = b$ and $\rho v^i = -\varepsilon^{ij}e_j$. However, it is crucial to note that this solution {\it solution} is not unique. We can define another gauge field $a'_{\mu}$ so that 
\begin{subequations}
\begin{align}
& \rho = b'\\
& \rho v^i = -\varepsilon^{ij}e'_j + \zeta\varepsilon^{ij}\p_j b'
\end{align}
\end{subequations}
and it still perfectly satisfies the conservation law equation because the boundary term $\varepsilon^{ij}\p_jb'$ has no divergence. For notational simplicity, we drop the prime, and the vorticity term given an arbitrary $\zeta$ is 
\begin{align}
\label{curlv}\varepsilon^{ij}\p_iv_j = \nabla_i\left( \frac{g^{ij}e_j}{b}\right) - \zeta\nabla_i\nabla^i\log b.
\end{align}
Plugging $\rho$ and $v^i$ back into~\eqref{Lmaster} rewrites it as a function of $\zeta$:
\begin{align}
& \mathscr L = \frac{m}{2b}g^{ij}e_i e_j - \varepsilon^{\mu\nu\rho}A_{\mu}\p_{\nu}a_{\rho} - s \varepsilon^{\mu\nu\rho}\omega_{\mu}^{\rm nc}\p_{\nu}a_{\rho} - \epsilon(b)\notag\\
& + \left( m\zeta - \frac{s}{2}\right) b\nabla_i \left( \frac{g^{ij}e_j}{b}\right) + \frac{\zeta}{2}(s-m\zeta)b\nabla_i\nabla^i\log b\notag\\
\label{Lsecondway}& -\zeta b (B + s\varepsilon^{ij}\p_i\omega_j),
\end{align}
where $\omega^{\rm nc}_{\mu}$, the non-covariant SO(2) connection, refers to Eq.~\eqref{omegaZero} without the vorticity term. 

To reconcile with the first derivation, we can choose $\zeta = s/(2m)$, which formally {\it eliminates} the Berry phase term $\nabla^i (e_i/b)$, yet suggests an improved connection by collecting terms linear in $b$: 
\begin{align}
\omega_0^{\rm nc} \to \omega_0^{\rm nc} + \frac{s}{2m}(B + s\varepsilon^{ij}\p_i\omega_j).
\end{align}
This identifies with the connection absorbed in $\bar v_0$ in~\eqref{v0bar}. The internal energy correction $\frac{s^2}{8m}b\nabla_i\nabla^i\log b$ has a precise correspondence in Eq.~\eqref{Lfirstway} as well. In light of these, we again conclude that the dual Lagrangian proposed in Ref.\cite{PhysRevB.91.064508} is a result of truncating $s^2$ terms in Eq.~\eqref{Lsecondway} evaluated at $\zeta = s/(2m)$. The Lagrangians in Ref.\cite{PhysRevB.89.174507} and Ref.\cite{PhysRevB.91.064508} are equivalent because they both stem from~\eqref{Lmaster} with the same amount of truncation.

Let us clarify that the choice of $\zeta$ does not have physical implicationsl it merely re-parametrizes $v^i$ by redefining the dual electric field $e_i$. Physical quantities, such as the U(1) current $ J^{\mu} = -g^{-1/2}{\delta S}/{\delta A_{\mu}}$, when expressed in terms of solutions to the field equations, do not depend on $\zeta$. Since the current is determined by $v^i$ and $\rho$ and $\rho=b$ is $\zeta$-independent, the curl ~\eqref{curlv} and the divergence of $v^i$ 
\begin{align}
\nabla_i v^i = -\varepsilon^{ij}\p_i\left( \frac{e_j}{b}\right)
\end{align}
also do not depend on $\zeta$ by Helmholtz decomposition. This statement can be verified using the Maxwell equations derived from~\eqref{Lsecondway}. Because this degree of freedom is not sensitive to spacetime curvature, we can work out the proof in the flat background $g_{ij} = \delta_{ij}$ without losing generality.

Varying~\eqref{Lsecondway} with respect to $a_0$ produces the Gauss law:
\begin{align}
\label{GaussLaw}\nabla\cdot\left( \frac{\mb e}{b}\right) = \frac{B}{m} + \left( \zeta - \frac{s}{2m}\right)\nabla^2\log b.
\end{align}
This immediately implies that the vorticity,
\begin{align}
\epsilon^{ij}\p_iv_j = \nabla\cdot\left( \frac{\mb e}{b}\right) - \zeta \nabla^2\log b = \frac{B}{m} - \frac{s}{2m}\nabla^2\log b,
\end{align}
is $\zeta$-independent. Variation with respect to $a_i$ yields the Amp\`{e}re's law:
\begin{align}
& -m\p_t \frac{e_i}{b} - \frac{m}{2}\epsilon^{ij}\p_j \left( \frac{\mb e}{b}\right)^2 + \epsilon^{ij}E_j - \epsilon^{ij}\epsilon''(b)\p_j b\notag\\
+ & \left(m\zeta - \frac{s}{2}\right) \left[ \epsilon^{ij}\p_j \nabla\cdot\frac{\mb e}{b} + \p_t\p_i \log b + \epsilon^{ij}\p_j \frac{\mb e\cdot\nabla\log b}{b}\right]\notag\\
+ & \frac{\zeta}{2}(s-m\zeta)\left[ 2\epsilon^{ij}\p_j \nabla^2\log b + \epsilon^{ij}\p_j (\nabla\log b)^2\right]\notag\\
\label{AmpereLaw}- & \zeta \epsilon^{ij}\p_j B = 0.
\end{align}
Taking the curl of the above equation,
\begin{align}
&-m\p_t \epsilon^{ij}\p_i \frac{e_j}{b} + \frac{m}{2}\nabla^2\left( \frac{\mb e}{b}\right)^2 - \nabla\cdot\mb E + \nabla\cdot(\epsilon''\nabla b)\notag\\
\label{AmpereCrossE}-&\left(m\zeta - \frac{s}{2}\right) \nabla^2\left[ \nabla\cdot\frac{\mb e}{b} + \frac{\nabla\log b\cdot\mb e}{b}\right]\\
- & \frac{\zeta}{2}(s-m\zeta)\nabla^2\left[2\nabla^2\log b+ (\nabla\log b)^2 \right] + \zeta\nabla^2B = 0.\notag
\end{align}
Using~\eqref{GaussLaw}, the divergence $\nabla\cdot(\mb e/b)$ is absorbed by $\nabla^2B$ and $\nabla^4\log b$, with coefficients:
\begin{align*}
&\nabla^2B:\, -\left(m\zeta - \frac{s}{2}\right)\frac{1}{m} + \zeta = \frac{s}{2m}\\
& \nabla^4\log b:\, -\left( m\zeta - \frac{s}{2}\right) \left( \zeta - \frac{s}{2m}\right) - \zeta(s-m\zeta) = -\frac{s^2}{4m}. 
\end{align*}
Hence, Eq.~\eqref{AmpereCrossE} can be reorganized as 
\begin{align}
& m\p_t\epsilon^{ij}\p_i \frac{e_j}{b} = -\nabla\cdot\mb E + \nabla\cdot(\epsilon''\nabla b) \notag\\
\label{reorgAmpere}- & \frac{s^2}{4m}\nabla^4\log b + \frac{s}{2m}\nabla^2B\\
+ & \frac{m}{2}\nabla^2\left\{\left[ \frac{\mb e}{b} - \zeta\nabla\log b\right]\cdot\left[\frac{\mb e}{b} -\left( \zeta - \frac{s}{m}\right)\nabla\log b \right]\right\}.\notag
\end{align}
The only $\zeta$ dependences are garnered in the last line and appear in terms of the combination $\frac{\mb e}{b} - \zeta\nabla\log b$. Again, by using~\eqref{GaussLaw}, this could be solved formally as:
\begin{align}
-\nabla\int d\mb x'\, G(\mb x, \mb x')\left( \frac{B(\mb x')}{m} - \frac{s}{2m}\nabla^2\log b(\mb x')\right),
\end{align}
where $G(\mb x, \mb x')$ is the Green's function of the 2-dimensional Poisson equation. As a result, the right-hand side of~\eqref{reorgAmpere} is also independent of $\zeta$. This completes the proof that $\p_iv^i$ and $\epsilon^{ij}\p_iv_j$, and thereby $v^i$, do not depend on $\zeta$.

To conclude this subsection, we reckon that the truncating the $\mathscr O(s^2)$ terms is legitimate within the context of a leading-order effective theory construction based on the power counting scheme for gradient expansion, where $\rho$, $A_{\mu}$ and $g_{ij}$ are all $\mathscr O(1)$; that is, the mass dimensions $[\rho] = [A_{\mu}] = [g_{ij}] = 0.$ In Eq.~\eqref{rotv}, we observe that $[B] = 1$ and $[\p_i\omega_j] = [\nabla^2\log\rho] = 2$. To maintain $v^i$ as $\mathscr O(1)$, it is consistent to disregard the fluctuation induced by $\nabla\log\rho$. Nevertheless, we will see soon that these next-to-leading fluctuations are non-trivial. Their possession of an extra factor of $m^{-1}$ becomes crucial when performing the lowest Landau level projection, where $m\to 0$.
\subsection{Electromagnetic response}
We end this section with the linear electromagnetic response predicted by Lagrangian~\eqref{Lsecondway} without truncation at $\zeta = 0$. This exercise not only showcases the effect of density fluctuations but also prepares the solution for the bosonic superfluid in the LLL to reference. Linearized around the mean-field value $(\lan b\ran, \lan e_i\ran) = (n_0, 0) = (\rho_0 m, 0)$, Eqs.~\eqref{GaussLaw} and~\eqref{AmpereLaw} read:
\begin{subequations}
\begin{align}
\label{linearGauss}\nabla\cdot\mb e&= \rho_0B - \frac{s}{2m}\nabla^2b\\ 
\p_t e_i &+ \rho_0\epsilon''\epsilon^{ij}\p_j b \notag\\
\label{linearAmpere}&+ \frac{s}{2m}\p_t\p_i b + \frac{s}{2m}\epsilon^{ij}\p_j \nabla\cdot\mb e = \rho_0\epsilon^{ij}E_j.
\end{align}
\end{subequations}
Plugging~\eqref{linearGauss} into~\eqref{linearAmpere}, taking the curl of the Amp\`{e}re's law, and substituting $\p_tb$ for $\epsilon^{ij}\p_ie_j$ using the Bianchi identity, we could solve for 
\begin{subequations}
\begin{align}
\label{bSolution}b = \rho_0 \frac{i\mb p\cdot\mb E + \frac{s}{2m}\mb p^2B}{\omega^2 -c_s^2\mb p^2}, c_s^2 = \rho_0\epsilon'' + \left(\frac{s}{2m}\right)^2\mb p^2,
\end{align}
where we have Fourier transformed the derivatives $(\p_t, \nabla)\to (-i\omega, i\mb p)$. Replacing the magnetic field in~\eqref{linearAmpere} with this solution, we deduce 
\begin{align}
\label{eSolution}&-\epsilon^{ij}e_j =  \frac{i\omega \rho_0}{\omega^2 - c_s^2\mb p^2}E_i\\
& + i \left( 1 - \frac{s^2\mb p^2}{4m^2c_s^2}\right) \frac{-c_s^2\rho_0}{\omega^2 - c_s^2\mb p^2}\epsilon^{ij}p_j B - \frac{s}{2m}\frac{\rho_0\mb p^2\epsilon^{ij}E_j}{\omega^2 - c_s^2\mb p^2}.\notag
\end{align}
\end{subequations}
The effect of the $s^2$ term manifests in the dispersion relation, introducing fluctuations into the wave speed $c_s^2$. Eqs.~\eqref{bSolution} and~\eqref{eSolution} constitute the U(1) current $J^{\mu} = (b, -\epsilon^{ij}e_j)$, consistent with the solution in Ref.\cite{PhysRevB.89.174507} in the limit of long wavelength.

To recap, in this section we clarified the next-to-leading contribution to the vorticity that has been omitted in the current literature on chiral $p$-wave superfluids. The context-dependent reasons for truncation and inclusion were discussed. Additionally, it was pointed out that in the dual-photon framework, there is a degree of freedom for separating the boundary current from the definition of the dual electric field, while the physical quantities do not depend on this choice. We will further demonstrate the importance of the next-to-leading term shortly in the sections to follow.
\section{Bosonic superfluid in the LLL}\label{swave}
\subsection{Effective Lagrangian from the master Lagrangian}
In this section, we establish the proclaimed formal correspondences between the scalar bosonic superfluid in the lowest Landau level \cite{PhysRevLett.122.235301} and the $p$-wave fermionic chiral superfluid presented in the preceding section. To that end, the effective Lagrangian shall be derived within the framework of non-relativistic diffeomorphism. As argued in Refs.\cite{Son:2013rqa, PhysRevLett.122.235301}, a simple bosonic version of~\eqref{Lnon} is inadequate for the sake of the lowest Landau-level projection $m\to 0$. A non-minimal coupling to the magnetic field, controlled by the $\mathsf g$-factor, is required:
\begin{align}
\mathscr L_{\mathsf g} = \mathscr L_{\rm non} + \varsigma \frac{\mathsf g B}{4m}\psi^{\dag}\psi.
\end{align}
The LLL limit is regular for $\mathsf g=\hat{\mathsf g}:=2$, and the model for general values of $\mathsf g$ is extrapolated via $\mathscr L_{\mathsf g}(\cdots, A_0) = \mathscr L_{\hat{\mathsf{g}}}(\cdots, A_0 + \frac{\mathsf g -\hat{\mathsf g}}{4m}B)$. To apply the diffeomorphism approach, equations~\eqref{dA0} and~\eqref{tildeA0} are generalized to
\begin{align}
 \delta A_0 = -\xi^k\p_k A_0 - A_k\dot{\xi}^k + \frac{\varsigma \mathsf g}{4}\varepsilon^{ij}(g_{jk}\dot{\xi}^k)
\end{align}
and
\begin{align}
\tilde{A}_0 = A_0 - \varsigma \frac{m}{2}g_{ij}v^iv^j - \varsigma \frac{ \mathsf g}{4}\varepsilon^{ij}\p_iv_j
\end{align}
respectively. In terms of the further improved $\tilde A_0$, the effective Lagrangian is again given by the master Lagrangian~\eqref{Lmaster} with $s = 0$. 

It is clear that $\mathsf g$ introduces the vorticity to the model, reminiscent of the role of $s$, and as a consequence, solving $v^i$ directly cannot reduce the Lagrangian into the exact Legendre transform of the energy density without truncation. Nevertheless, we could still rewrite $\rho$ and $v^i$ using a dual photon field by integrating $\theta$ in~\eqref{Lmaster}.  Choosing $\zeta = 0$ and recalling~\eqref{curlv}, we obtain
\begin{align}
\label{Lzeta}\mathscr L_{\mathsf g} = & \frac{mg^{ij}e_ie_j}{2b}  - \epsilon(b) \notag\\
& + \varsigma  \varepsilon^{\mu\nu\rho}A_{\mu}\p_{\nu}a_{\rho} - \frac{\mathsf g}{4}b\nabla_i\left( \frac{g^{ij}e_j}{b}\right),
\end{align}
which is identical to~\eqref{Lsecondway} with an replacement $s\to  \mathsf g/2$ and $\varsigma = -1$. The improved gauge potential is also introduced in Ref.\cite{PhysRevLett.122.235301} to derive the above effective Lagrangian. That derivation exploits the insight that the leading contribution in gradient expansion is simply $\varepsilon^{\mu\nu\rho}\tilde{A}_{\mu} \p_{\nu}a_{\rho}$, which, as we clarify in this work, stems from a choice of $\zeta$.

The full effective theory comprises the superfluid at $\mathsf g = \hat{\mathsf g} = 2$, the extrapolation term $(\mathsf g-\hat{\mathsf g})Bb/(4m)$, and the vortex crystal sector, which we adopt directly from Ref.\cite{PhysRevLett.122.235301} and Ref.\cite{10.21468/SciPostPhys.5.4.039}. The total Lagrangian is given by:
\begin{align}
\mathscr L_{\rm tot} = & \mathscr L_{\hat{\mathsf g}} + \frac{\mathsf g-\hat{\mathsf g}}{4m}Bb \notag\\
\label{Ltot}& - \frac{B_0}{2}b\epsilon^{ij}u^iD_tu^j + B_0e_i u^i - \mathscr E_{\rm el}(u_{ij}).
\end{align}
$u^i(x)$ is the field representing the vortex displacement. For the remainder of this section, we will set $\varsigma = 1$ so that the charge convention remains consistent with the adopted model.  To distinguish the additional source from $B_0$, we define the probe field $\hat A_{\mu}$ as 
\begin{subequations}
\begin{align}
& \hat A_0 = A_0 + \frac{(\mathsf g - \hat{\mathsf g})B_0}{4m}\\
& \hat B = B - B_0 = \varepsilon^{ij}\p_i \hat A_j.
\end{align}
\end{subequations}
The Lagrangian \eqref{Ltot} is modified as follows:
\begin{align*}
\mathscr L_{\hat{\mathsf g}}(A) + \frac{\mathsf g-\hat{\mathsf g}}{4m}Bb \to \mathscr L_{\hat{\mathsf g}}(\hat A)+ \frac{(\mathsf g - \hat{\mathsf g})}{4m}\hat Bb,
\end{align*} 
up to irrelevant constants.
\subsection{Linearized model as a gauge theory and its solution by analogue}
Having elucidated the correspondence between the superfluid sectors, we now turn our attention to the correspondence between the linear response calculations. 

We first examine the collective motion of the vortex crystal. In the background field configuration where $B = B_0$ and the rest of $A_{\mu}$ vanishes (i.e., $\hat A = 0$), the linearized Lagrangian reads:
\begin{subequations}
\begin{align}
\label{LvcLinear}\mathscr L_{\rm lin} = & \frac{m\mb e^2}{2n_0} + \frac{\hat{\mathsf g}}{4n_0}e^i\p_i b - \frac{\epsilon'' b^2}{2} + B_0 e_i u^i \notag\\
&- \frac{B_0n_0}{2}\epsilon_{ij}u^i\dot{u}^j - \mathscr E^{(2)}_{\rm el}(u_{ij}).
\end{align}
Examining the dual photon sector of the Lagrangian, we observe that by defining an additional photon $\hat{a}_{\mu} = (0, B_0\epsilon_{ij}u^j)$, the term $B_0 e_i u^i$  becomes $-\epsilon^{\mu\nu\rho}a_{\mu}\p_{\nu}\hat a_{\rho}$ up to a boundary term. The associated auxiliary electric and magnetic fields are $\hat e_i = B_0\epsilon_{ij}\p_tu^j$ and $\hat b = -B_0\nabla\cdot\mb u$. Substituting these into the expression, Eq.~\eqref{LvcLinear} becomes
\begin{align}
\mathscr L_{\rm lin} = & \frac{m\mb e^2}{2n_0} + \frac{\hat{\mathsf g}}{4n_0}e^i\p_ib -\frac{\epsilon''}{2}b^2 - \epsilon^{\mu\nu\rho}a_{\mu}\p_{\nu}\hat{a}_{\rho} \notag\\
\label{LvcLinearAhat}& + \frac{n_0}{2B_0}\epsilon^{\mu\nu\rho}\hat a_{\mu}\p_{\nu}\hat a_{\rho} - \mathscr E^{(2)}_{\rm el}(\hat b, \p_t^{-1}\nabla\cdot\hat{\mb e}).
\end{align}
\end{subequations}
In the above, we have utilized the fact that the derivatives $\p u$ in $\mathscr E^{(2)}$ can be rewritten as combinations of $\nabla\times\mb u$ and $\nabla\cdot\mb u$, which are proportional to $\p_t^{-1}\nabla\cdot\hat{\mb e}$ and $\hat b= \p_t^{-1}\nabla\times\hat{\mb e},$ respectively. In terms of $\hat{a}_{\mu}$, we can immediately recognize that the dual photon sector of Eq.~\eqref{LvcLinear} is identical to the linearized version of Eq.~\eqref{Lsecondway} in the flat background $g_{ij} = \delta_{ij}$, with the replacement $A_{\mu}\to \hat{a}_{\mu}$, $s\to \hat{\mathsf g}/2$, and the choice $\zeta = 0$. By analogue to~\eqref{bSolution} and~\eqref{eSolution}, the dual photon field can be solved as:
\begin{subequations}
\begin{align}
\frac{b}{\rho_0 B_0} &=\frac{1}{B_0}\frac{i\mb p\cdot\hat{\mb e} + \frac{\hat{\mathsf g}}{4m}\mb p^2\hat{b}}{\omega^2 - c_s^2\mb p^2}, \notag\\
&= \frac{\left(\omega \epsilon_{ij}p_iu^j - \frac{\hat{\mathsf g}}{4m}\mb p^2i\mb p\cdot\mb u\right)}{\omega^2 - c_s^2\mb p^2}\\
\frac{e_i}{\rho_0 B_0} 
&= \frac{\left[i\omega \epsilon_{ij}\hat e_j+ i \rho_0\epsilon'' p_i\hat b + \frac{\hat{\mathsf g}\mb p^2}{4m}\hat e_i \right]}{B_0(\omega^2- c_s^2\mb p^2)}\notag\\
& = \frac{\left[-\omega^2u^i + \rho_0\epsilon''p_i \mb p\cdot\mb u-i \frac{\hat{\mathsf g}\omega\mb p^2}{4m}\epsilon_{ij}u^j\right]}{\omega^2- c_s^2\mb p^2}.
\end{align}
\end{subequations}
From these solutions, it becomes apparent that the fluctuation term in $c_s^2$ is relevant in the lowest Landau-level limit $m\to 0$, where the propagator reduces to:
\begin{align}
\frac{1}{\omega^2 - c_s^2\mb p^2}\to -\frac{16m^2}{\hat{\mathsf g}^2\mb p^4}.
\end{align}
Thereby, we can expand the dual photon solutions as:
\begin{subequations}
\begin{align}
\label{bInAhat}\frac{b}{B_0}\to & -\frac{1}{B_0}\frac{4n_0}{\hat{\mathsf g}\mb p^2}\hat b = \frac{4n_0}{\hat{\mathsf g}\mb p^2}i \mb p\cdot\mb u,\\
\label{eInAhat}\frac{e_i}{B_0}\to & -\frac{1}{B_0}\left(\frac{4n_0}{\hat{\mathsf g}\mb p^2}\hat e_i + \frac{16n_0^2\epsilon''}{\hat{\mathsf g}^2\mb p^4}i p_i\hat b\right)\notag\\
= & \frac{4n_0i\omega\epsilon_{ij}u^j}{\hat{\mathsf g}\mb p^2} - \frac{16n_0^2\epsilon''}{\hat{\mathsf g}^2\mb p^4}p_i \mb p\cdot\mb u.
\end{align}
\end{subequations}
To proceed, let us include the vortex crystal sector by considering the explicit elastic energy \cite{PhysRevLett.122.235301, 10.21468/SciPostPhys.5.4.039, 10.21468/SciPostPhys.9.5.076, nguyen2023quantum, Sonin_2016, RevModPhys.59.87}, up to boundary terms:
\begin{align}
\label{elasticE}\mathscr E_{\rm el}^{(2)} = 2C_1(\nabla\cdot\mb u)^2 + C_2[(\nabla\cdot\mb u)^2 + (\nabla\times \mb u)^2].
\end{align}
Together with Eq.~\eqref{LvcLinear}, this leads to
\begin{subequations}
\begin{align}
&-B_0n_0\epsilon_{ij}\p_tu^j + B_0 e_i \notag\\
\label{uEoM}&+ 2[(2C_1 + C_2)\p_i (\nabla\cdot\mb u) - C_2\epsilon_{ij}\p_j(\nabla\times\mb u)] = 0.
\end{align}
The second line can be simplified as: $2[2C_1\p_i(\nabla\cdot\mb u) + C_2\nabla^2u_i]$, but it is more convenient to work with the divergence and the curl of the field to switch between $u^i$ and $\hat a_{\mu}$ representations. Substituting $\hat{\mb e}$ and $\hat b$ for $\p u$, the equations of motion~\eqref{uEoM} read:
\begin{align}
&-n_0\hat e_i + B_0e_i \notag\\
\label{ehatEoM}& - \frac{2}{B_0}[(2C_1 + C_2)\p_i\hat b + C_2\epsilon_{ij}\p_j \p^{-1}_t\nabla\cdot\hat{\mb e}] = 0.
\end{align}
\end{subequations}
Combined with Eqs.~\eqref{bInAhat} and~\eqref{eInAhat}, we find that $\hat b$ and $\nabla\cdot\hat{\mb e}$ satisfy simple equations in the momentum representation $(\p_t, \nabla)\to (-i\omega, i\mb p)$. In the long-wavelength limit as $\mb p\to 0$, we have:
\begin{align}
\begin{pmatrix}
in_0\left( 1+ \frac{4B_0}{\hat{\mathsf g}\mb p^2} \right)& -\frac{16n_0^2\epsilon''B_0}{\hat{\mathsf g}^2\mb p^2} \\ 
-\frac{2iC_2\mb p^2}{B_0} & n_0 \omega^2\left( 1+ \frac{4B_0}{\hat{\mathsf g}\mb p^2}\right)
\end{pmatrix}\begin{pmatrix} \mb p\cdot\hat{\mb e} \\ \hat b\end{pmatrix} =0.
\end{align}
The characteristic equation gives rise to the Tkachenko modes:
\begin{align}
\label{mode}\omega^2 = \frac{2C_2\epsilon''\mb p^4}{B_0^2} - \frac{C_2\hat{\mathsf g}\epsilon''\mb p^6}{B_0^3} + \cdots.
\end{align}
With $\hat{\mb e}$ and $\hat b$, it is evident that contribution from $2C_1+C_2$ does not appear in the leading order of the collective excitation. From Eq.~\eqref{elasticE}, the term $(2C_1 +C_2)$ is associated with the {\it magnetic energy} $(\nabla\cdot\mb u)^2\sim \hat b^2$, which is reduced by a factor of $\mb p^4$ when translated to the magnetic energy of the dual photon by virtue of Eq.~\eqref{bInAhat}. Therefore, it becomes negligible compared to the internal energy curvature $\epsilon''$ in the two leading contributions.

We can now move toward the full electromagnetic response given by model~\eqref{Ltot} by turning on a background gauge field on top of $B_0$. The counterpart of Lagrangian~\eqref{LvcLinearAhat} reads:
\begin{align}
\label{LvcLinearAhatA}\mathscr L_{\rm lin} + \epsilon^{\mu\nu\rho}\hat{A}_{\mu}\p_{\nu}a_{\rho} + \frac{(\mathsf g - \hat{\mathsf g})}{4m}\hat Bb.
\end{align}
We aim to provide a similarly intuitive approach to extract the leading-order effective Lagrangian using $\hat a_{\mu}$ in the LLL limit $m\to0$. The effective Lagrangian as a functional of $\hat A_{\mu}$ is computed by integrating out $\hat a_{\mu}$ and $a_{\mu}$. This task can be significantly simplified by organizing the expected form of the gauge-invariant terms, which, in the leading order, should assume the following structure: \cite{PhysRevResearch.3.013103}
\begin{align}
\label{LLeading}\frac{\nu}{4\pi}\epsilon^{\mu\nu\rho}\hat{A}_{\mu}\p_{\nu}\hat{A}_{\rho} + \frac{\epsilon}{2}\hat{\mb E}^2 -\frac{1}{2\mu}\hat{B}^2+ \alpha \hat{\mb E}\cdot\nabla \hat{B} + \cdots.
\end{align}
The coefficients $\nu$, $\epsilon$, $\mu$, and $\alpha$ are determined by expanding solutions in small momenta and $m$. Additionally, we note that Eq.~\eqref{LvcLinearAhatA} remains valid because the vortex crystal sector does not depend on either $\hat A$ or $m$. However, solutions~\eqref{bInAhat} and~\eqref{eInAhat} are unfortunately not applicable, although solutions involving $\hat A$ can also be deduced by analogue. By counting the number of derivatives $\p_i$ in Eq.~\eqref{LvcLinearAhatA}, we find that it suffices to solve Eq.~\eqref{ehatEoM} to the order:
\begin{subequations}
\begin{align}
\label{eAndehat} e_i &= \frac{n_0}{B_0}\hat e_i + \mathscr O(C_I\mb p^2)\\
\label{bAndbhat} b &= \frac{n_0}{B_0}\hat b+ \mathscr O(C_I\mb p^2), I = 1, 2.
\end{align}
\end{subequations}
At this point, we conclude that the leading-order effective Lagrangian~\eqref{LLeading} does not involve $C_I$ without needing to fully engage in the remaining exercises. Denoting $\sigma_0 = \frac{n_0}{B_0}$,~\eqref{LvcLinearAhatA} simplifies to the following gauge theory:
\begin{align}
\label{LeffAhat}\mathscr L = &\frac{m\sigma^2_0}{2n_0}\hat{\mb e}^2 + \frac{\hat{\mathsf g}\sigma_0^2}{4n_0}\hat e^i\p_i\hat b - \frac{\epsilon''}{2}\sigma^2_0\hat b^2 - \frac{\sigma_0}{2}\epsilon^{\mu\nu\rho}\hat a_{\mu}\p_{\nu}\hat{a}_{\rho}\notag\\
&+ \frac{(\mathsf g-\hat{\mathsf g})}{4m}\hat B\sigma_0\hat b + \sigma_0\epsilon^{\mu\nu\rho}\hat A_{\mu}\p_{\nu}\hat{a}_{\rho}.
\end{align}
At small momenta, the physics is dominated by the Chern-Simons terms $\sigma_0\epsilon^{\mu\nu\rho}\hat A_{\mu}\p_{\nu}\hat{a}_{\rho} - \frac{\sigma_0}{2}\epsilon^{\mu\nu\rho}\hat a_{\mu}\p_{\nu}\hat{a}_{\rho}$, whose effective Lagrangian can be read off as $\frac{\sigma}{2}\epsilon^{\mu\nu\rho}\hat A_{\mu}\p_{\nu}\hat A_{\nu}$. 
To extract the other coefficients in Eq.~\eqref{LLeading}, we again refer to Eq.~\eqref{Lsecondway} to obtain the full field equations from Lagrangian~\eqref{LeffAhat}. These equations are analogous to~\eqref{linearAmpere} with some minor adjustments due to the presence of $\sigma_0$ and the Chern-Simons dynamics: 
\begin{subequations}
\begin{align}
&\sigma_0 \p_t \hat e_i + \epsilon''\sigma_0\rho_0\epsilon^{ij}\p_j \hat b-\rho_0\epsilon^{ij}\hat e_j + \frac{\hat{\mathsf g}}{4m}\sigma_0\p_t\p_i\hat b\notag\\
\label{hatEquation}& + \frac{\hat{\mathsf g}}{4m}\sigma_0\epsilon^{ij}\p_j\nabla\cdot\hat{\mb e}  = \frac{\mathsf g-\hat{\mathsf g}}{4m}\rho_0\epsilon^{ij}\p_j\hat B - \rho_0\epsilon^{ij}\hat E_j.
\end{align}
There is no Gauss law in the conventional sense because $\hat a_0 =0$ by construction. Nevertheless, the above equations are equivalent to two equations governing $\hat b\sim \nabla\times\hat{\mb e}$ and $\nabla\cdot\hat{\mb e}$. The counterpart of the Gauss law~\eqref{linearGauss} is obtained by taking the divergence of the above equation:
\begin{align}
\label{hatGauss}\sigma_0 \p_t\nabla\cdot\hat{\mb e} + \left[ \frac{\hat{\mathsf g}}{4m}\sigma_0\nabla^2 - \rho_0\right]\p_t\hat b = -\rho_0\p_t\hat{B}.
\end{align}
\end{subequations}
For a real gauge theory, this equation would be equivalent to the Gauss law, and therefore is redundant. This equation differs from previous Gauss law in that the Chern-Simons dynamics introduces the cyclotron frequency $\omega_c = B_0/m$ as a natural scale. The lowest Landau level limit $m\to 0$ is equivalent to $\omega_c$ with $B_0$ held fixed. 

To further utilize the solution by analogue technique, let us introduce the following quantities:
\begin{subequations}
\begin{align}
& \hat{c}_s^2 = \rho_0\epsilon'' + \frac{\hat{\mathsf g}\omega_c}{2m}\mb p^2 + \left(\frac{\hat{\mathsf g}\mb p^2}{4m}\right)^2\\
& \hat{\omega}_c = \omega_c + \frac{\hat{\mathsf g}}{4m}\mb p^2\\
& \Pi = \frac{\omega_c}{\omega^2 - \omega_c^2 - \hat{c}_s^2\mb p^2}.
\end{align}
\end{subequations}
In terms of these quantities, it can be verified that the magnetic field solution reads:
\begin{subequations}
\begin{align}
\label{bhatSolution}\hat b = -\Pi\left( {[\hat{\omega}_c + \frac{\mathsf g-\hat{\mathsf g}}{4m}\mb p^2]\hat B + i \mb p\cdot\hat{\mb E}}\right).
\end{align}
Comparing this to Eq.~\eqref{bSolution}, we see $\Pi$ generalizes the original propagator $\rho_0/(\omega^2 - c_s^2\mb p^2)$. Due to the extra $\sigma_0$ on the left-hand side of~\eqref{hatEquation}, the ground state superfluid density is rescaled by $1/\sigma_0$: $\rho_0\to\rho_0/\sigma_0 = \omega_c$. The corrections to the dispersion relations in both Eqs.~\eqref{bSolution} and~\eqref{bhatSolution} are given by the coefficients of the dynamical magnetic field in the associated Gauss laws~\eqref{linearGauss} and~\eqref{hatGauss}. The electric field dependence in Eq.~\eqref{hatEquation} is identical to that in Eq.~\eqref{linearAmpere}, up to a change in the sign of the charge, which implies the presence of the term $-i\mb p\cdot\hat{\mb E}$. Lastly, in Eq.~\eqref{bSolution}, the magnetic field dependence arise from its coupling to the Gauss law. Similarly, in Eq.~\eqref{hatEquation}, this coefficient is obtained by replacing $s$ with $\sigma_0\times\hat{\mathsf g}/2$, and it additionally receives a dispersionless contribution from the Chern-Simons dynamics, leading to: $\rho_0 - \frac{\hat{\mathsf g}}{4m}\sigma_0\nabla^2 = \sigma_0 \hat{\omega}_c$. The other coefficient directly inherits from the right-hand side of Eq.~\eqref{hatEquation}.

It is more subtle to accurately read off the electric field solution $\hat{e}_i$ using this approach compared to directly solving the equation. However, we can still observe that a significant portion of it mirrors Eq.~\eqref{eSolution}:
\begin{align}
-\frac{\epsilon^{ij}\hat e_j}{\Pi} =& -i\omega\hat{E}_i + \left( \rho_0\epsilon''-\frac{\mathsf g-\hat{\mathsf g}}{4m}\hat{\omega}_c\right)i\epsilon^{ij}p_j\hat{B} \notag\\
\label{ehatSolution}&+ \hat{\omega}_c\epsilon^{ij}\hat E_j + \frac{\mathsf g-\hat{\mathsf g}}{4m}p_i(\mb p\times\hat{\mb E}).
\end{align}
\end{subequations}
By replacing $s\mb p^2/(2m)$ and the propagator with $\hat{\omega}_c$ and $\Pi$, respectively, the Hall and longitudinal current can be constructed. To extract the coefficient of the Meissner term $i\epsilon^{ij}p_j\hat B$, we need to interpret the parenthesis in~\eqref{eSolution} as:
\begin{align*}
\frac{1}{\mb p^2}\left( c_s^2\mb p^2 - \left( \frac{s\mb p^2}{2m}\right)^2\right),
\end{align*}
where the $s^2$ results from the coefficient of $\nabla\cdot\mb e$ in Eq.~\eqref{linearAmpere} and $B$ in Eq.~\eqref{bSolution}. Plugging their counterparts into the current equation gives the coefficient of $i\epsilon^{ij}p_j\hat B$: 
\begin{align*}
\left[(\omega_c^2 + \hat c_s\mb p^2) - \hat{\omega}_c(\hat{\omega}_c + \frac{\mathsf g-\hat{\mathsf g}}{4m}\mb p^2)\right] =\left(\epsilon''\rho_0 -\frac{\mathsf g-\hat{\mathsf g}}{4m}\hat{\omega}_c\right)\mb p^2.
\end{align*}
The last term in Eq.~\eqref{ehatSolution}, proportional to $p_i(\mb p\times\hat{\mb E})$, is generated combining the Chern-Simons dynamics with Eq.~\eqref{bhatSolution} and does not have an equally transparent correspondence in the $p$-wave chiral superfluid problem.

Solutions~\eqref{bhatSolution} and~\eqref{ehatSolution} completely determine the linear approximation of the model~\eqref{LeffAhat}. Though superficially intricate, they can be organized in a double expansion in $p_i$ and $\omega_c^{-1}$, resulting in interpretable field identifications similar to Eqs.~\eqref{eAndehat} and~\eqref{bAndbhat}. We illustrate the expansion by constructing the U(1) response current: $J^i = \sigma_0 (-\epsilon^{ij}\hat e_j + \frac{\mathsf g-\hat{\mathsf g}}{4m}\epsilon^{ij}\p_j\hat b)$. Using 
\begin{align}
\Pi = -\frac{1}{\omega_c} + \left[ \frac{\hat{\mathsf g}}{2\omega_cB_0} + \frac{\sigma_0\epsilon''}{\omega_c^2}\right]\mb p^2 + \cdots,
\end{align}
the leading order constituents of $\hat b$ are given by
\begin{subequations}
\begin{align}
\label{bhatExpansion}\hat b = \hat B -\frac{\mathsf g-2\hat{\mathsf g}}{4B_0}\nabla^2\hat B + \frac{1}{\omega_c}\nabla\cdot\hat{\mb E} + \cdots.
\end{align}
The omitted terms $(\cdots)$ consist of higher order corrections of $\mathscr O(\nabla)$ or $\mathscr O(\omega_c^{-1})$. Similarly, the electric field to $\mathscr O(\omega_c^0)$ can be expressed as:
\begin{align}
\label{ehatExpansion}-\epsilon^{ij}\hat e_j =& -\epsilon^{ij}\hat{E}_j - \frac{\hat{\mathsf g}}{4B_0}\nabla^2\epsilon^{ij}\hat{E}_j + \frac{\mathsf g-\hat{\mathsf g}}{4B_0}\p_i(\nabla\times\hat{\mb E})\notag\\
&+ \left( -\sigma_0\epsilon'' + \frac{\mathsf g-\hat{\mathsf g}}{4B_0}\omega_c\right) \epsilon^{ij}\p_j\hat B + \cdots.
\end{align}
\end{subequations}
Together, these imply the total current:
\begin{align}
J^i = & - \sigma_0\left( 1 -\frac{\mathsf g-2\hat{\mathsf g}}{4B_0}\nabla^2\right) \epsilon^{ij}\hat E_j\notag\\
\label{Jexpansion}&  + \sigma_0\left( \frac{\mathsf g-\hat{\mathsf g}}{2B_0}\omega_c - \sigma_0\epsilon'' \right)\epsilon^{ij}\p_j\hat{B}+ \cdots.
\end{align}
The conductivity tensor is defined as $\sigma^{ij} = -\frac{\delta J^i}{\delta E^j}$. The Hall and longitudinal components are given by $\sigma_H = \frac{1}{2}\epsilon_{ji}\sigma^{ij}$ and $\sigma_L = \frac{1}{2}\delta_{ij}\sigma^{ij}$, respectively. From this expansion, we can identify the origins of the Hall response: $\sigma_0$ from the Chern-Simons term, $\frac{-\hat{\mathsf g}}{4B_0}\mb p^2\sigma_0$ from the Berry phase term (the same as the chiral $p$-wave superfluid in Eq.~\eqref{eSolution}), and $(\mathsf g-\hat{\mathsf g})\mb p^2\sigma_0/(4B_0)$ from the extrapolation term $\frac{\mathsf g-\hat{\mathsf g}}{4m}\hat{B}\hat b$. 

The effective action can be evaluated by plugging the solutions back into Lagrangian~\eqref{LeffAhat}. However, it will require properly expanding $\hat{\mb e}$ to $\mathscr O(\omega_c^{-1})$. Should one carry out this exercise, it becomes apparent that $\hat{\mb e}^2, \epsilon^{\mu\nu\rho}\hat a_{\mu}\p_{\nu}\hat{a}_{\rho}$, and $\epsilon^{\mu\nu\rho}\hat A_{\mu}\p_{\nu}\hat a_{\nu}$ all produce parity-odd contributions $\propto \hat B\nabla\cdot\hat{\mb E}$ in the Lagrangian. Remarkably, these contributions compensate for one another, leaving the residual parity-odd response attributable to the physics highlighted in the analysis above. A more enlightening approach is to integrate the relation:
\begin{align}
\delta\mathscr L_{\rm eff} = \delta\hat{A}_{\mu}J^{\mu} = \delta \hat{A}_0\sigma_0\hat b +\delta\hat A_i J^i,
\end{align}
where $\hat b$ and $J^i$ have been expanded to adequate orders in Eqs.~\eqref{bhatExpansion} and~\eqref{Jexpansion}. Up to $\mathscr O(\omega_c^0)$ and $\mathscr O(p^3)$,
\begin{align}
\mathscr L_{\rm eff} = &\frac{\sigma_0}{2}\epsilon^{\mu\nu\rho}\hat A_{\mu}\p_{\nu}\hat A_{\rho}  + \left(\frac{\mathsf g-\hat{\mathsf g}}{4B_0}\omega_c\sigma_0 - \frac{\sigma_0^2\epsilon''}{2} \right) \hat B^2\notag\\
& + \frac{\sigma_0}{4B_0}(\mathsf g-2\hat{\mathsf g})\hat B\nabla\cdot\hat{\mb E}.
\end{align} 
It is worth noting that although the field $\hat a_{\mu}$ has a vanishing temporal component (i.e., ${\hat a}_0 = 0$), the solutions exhibited in this section are automatically gauge-invariant. This is because the equations were solved and expressed purely in terms of field strengths. For the not manifestly invariant term $\epsilon^{\mu\nu\rho}\hat{a}_{\mu}\p_{\nu}\hat{a}_{\rho} = -\epsilon^{ij}\hat a_i\hat e_j$, we can also confirm gauge invariance by formally defining $\hat a_i(t) = \int^td\tau\, \hat e_{i}(\tau)$. Note that for our auxiliary electromagnetic field, it is the definition of $\hat{a}_i$ rather than a gauge choice. Each term in the expansion naturally organizes itself into a gauge invariant structure. For instance,
\begin{align*}
\epsilon^{ij}\int^td\tau\hat E_i(\tau)\hat E_j(t) = [ \epsilon^{ij}\hat A_i\hat E_j - \hat A_0\hat B] = -\epsilon^{\mu\nu\rho}\hat A_{\mu}\p_{\nu}\hat A_{\rho}.
\end{align*}
To summarize, in this section, we have developed the superfluid sector of the bosonic superfluid in a strong magnetic field $B_0$, utilizing non-relativistic diffeomorphism. We demonstrated that, in  flat spacetime, the Berry phase term can be identified with the vorticity term in Sec.\ref{pwave}. In the linear regime, the model can be formally expressed as a gauge theory by redefining the vortex crystal field as a photon $\hat a$. As a result, the problem mirrors a that of superfluid in an external electromagnetic probe, allowing the response functions to be extracted through analogy.

Before moving forward, we point out that our solution by analogue is limited within electromagnetic computations. The $s\to \mathsf g/2$ replacement, nevertheless, does not provide conclusions for gravitational responses such as the Hall viscosity. It is intriguing that, from the flat space point of view, $s$ produces a physical effect that $\hat{\mathsf g}$ does not, even though they together form the coefficient of the Berry phase. Under this circumstance, constructing the effective action in a general curved background is advantageous, as $s$ manifests itself as the Wen-Zee term.

\section{concluding remarks}\label{conclusion}
In summary, we dedicated effort to bridging the models of the $p$-wave chiral superfluid and the bosonic superfluid in the LLL. It is shown that by incorporating a correction to the existing description of the former, the solutions to the latter can be encapsulated in a simpler, organized and covariant manner. In addition to polishing the existing framework and elucidating the structure of the Berry phase for the $p$-wave chiral superfluid, it is enlightening to highlight that next-to-leading contributions in certain context could be instrumental in another, specifically the LLL projection in this context. The reformulation of elastic theory in terms of a non-local electromagnetism also supplies an alternative perspective on this classic problem. Our approach is yet another exhibition of the versatility of the method of effective theory, wherein similar or identical universal physics arises in various scenarios, allowing us to endow the same equations with numerous interpretations and perspectives. 

While most efforts have been devoted to resolving the theoretical connection between the two models, it is worthwhile to consider potential experimental implications. For the $p$-wave superfluid, the model presented in this work has often been treated as a toy model for the A phase of spin-polarized superfluid ${}^3$He. The correction proposed here suggests that by measuring the dispersion $\omega^2$ of the Goldstone mode, the parameter $s^2$ can be extracted at the order of $\mb p^4$, up to a factor involving the inverse effective mass of the underlying fermion. Meanwhile, the $s$-wave superfluid in the LLL has already been realized in cold-atom experiments \cite{Zwierlein2005, fletcher2021geometric, Mukherjee2022}. The effective action~\eqref{mainLeff} should govern all linear electromagnetic responses, particularly parity-odd responses such as the Hall conductivity. Additionally, our framework predicts a Hall viscosity of zero from symmetry and geometry considerations\cite{PhysRevLett.122.235301}, in contrast to the prediction of Ref.\cite{PhysRevB.110.024515}. An experimental determination of this response property would provide valuable insights into the proper construction of effective theories.

Let us comment on a few natural follow-up problems to this project. A key question is whether a $p$-wave chiral superfluid in the LLL can be explored in the same manner. To answer this question, we need to extract the conceptual essence of placing an effective theory in the LLL in the current framework. Firstly, we require the knowledge about extrapolating effective Lagrangians characterized by different $\mathsf g$ factors, as the LLL limit is only regular when $\mathsf g = 2$. This mechanism is elucidated in Sec.\ref{swave}. In addition, we must incorporate the vortex degrees of freedom into the full effective Lagrangian due to finite vorticity sourced by a background magnetic field (c.f. Eq.~\eqref{rotv}). For an $s$-wave bosonic superfluid, the dominant physics, parametrized by the $u^i$ field, is known from various studies \cite{PhysRevLett.110.181601, 10.21468/SciPostPhys.5.4.039}, which form part of the basis of this paper. However, for a $p$-wave chiral superfluid, there are more terms permitted up to the next-to-leading order in the effective Lagrangian, and the resulting dual electromagnetism is expected to be more intricate, thus invalidating the elegant solution by analogy proposed here. We refer readers to Ref.\cite{PhysRevB.91.064508, PhysRevB.92.035136} for relevant attempts. 

Another conceptually interesting question concerns the role of the auxiliary field $\hat a_{\mu}$ in other recent reformulations of the vortex crystal \cite{10.21468/SciPostPhys.9.5.076, nguyen2023quantum}. For instance,  in Ref.\cite{nguyen2023quantum}, it is shown that $u^i$ has only one degree of freedom in the leading order of the effective theory, given by $u^i = \epsilon^{ij}\p_j\phi/B_0$. Consequently, the vortex crystal can be viewed as a Lifshitz model in disguise. In terms of this parametrization, $\hat a_{i} = -\p_i\phi$, which appears to be a pure gauge. A physical interpretation of the corresponding gauge theory would be desirable. 

In addition to these, several questions arise from nuanced ingredients that could be relevant in a more realistic model of chiral superfluids. Of particular interest, from a dynamical perspective, is the existence of Coulomb interaction, which can change the nature of collective motions and generalize the effective Lagrangian to the realm of superconductors \cite{PhysRev.159.330, PhysRevB.77.144516, 10.21468/SciPostPhys.9.1.006, Hsiao_2023}. This should be coherently formulated using the gauge theory framework uncovered in Sec.\ref{swave}. From a geometric point of view, it is intriguing to explore the effect of and the structure of the Berry phase contribution from intrinsic geometric objects--sometimes dubbed the emergent gravity--that arise in $p$-wave chiral superfluids \cite{PhysRevB.98.064503, PhysRevB.100.094510, PhysRevB.100.104512}. Last but not least, the development of EFT with realistic microscopic spin can be crucial for accurately describing realistic chiral superfluids, such as ${}^3$He-A \cite{PhysRevB.103.064506}.

\begin{acknowledgments}
We thank Sergej Moroz for enlightening discussion on this topic.
\end{acknowledgments}

\bibliography{citation}
\end{document}